\documentclass[twocolumn,aps,prl,floatfix,superscriptaddress]{revtex4-1}
\usepackage{amsmath,amssymb,amsfonts}
\usepackage{mathrsfs}
\usepackage{graphicx,float}
\usepackage{times,txfonts}
\usepackage[makeroom]{cancel}
\usepackage{color}
\usepackage{multirow}
\usepackage{bbold}
\usepackage{color}
\usepackage{soul}
\usepackage{hyperref}
\usepackage{times,txfonts}
\usepackage{natbib}
\newcommand{\abs}[1]{\left\vert#1\right\vert}
\newcommand{\ass}{\stackrel{\textup{\tiny def}}{=}}

\newcommand{\cv}[1]{\Delta_{#1}}

\newcommand{\ket}[1]{\left| #1 \right\rangle}
\newcommand{\braket}[1]{\langle #1 \rangle}

\newcommand{\etal}{\textit{et~al.}}
\renewcommand{\epsilon}{\varepsilon}

\def\VR{\kern-\arraycolsep\strut\vrule &\kern-\arraycolsep}
\def\vr{\kern-\arraycolsep & \kern-\arraycolsep}
\usepackage{sistyle}
\usepackage{soul}
\sethlcolor{lightblue}

\begin{document}
\title{Experimental Tests of Multiplicative Bell Inequalities}
\author{Dilip Paneru}
\affiliation{Department of Physics, University of Ottawa, 25 Templeton Street, Ottawa, Ontario, K1N 6N5 Canada}
\author{Amit Te'eni}
\affiliation{Faculty of Engineering and the Institute of Nanotechnology and Advanced Materials, Bar Ilan University, Ramat Gan 5290002, Israel}
%%%
\author{Bar Y. Peled}
\affiliation{Center for Quantum Information Science and Technology \& Faculty of Engineering Sciences, Ben-Gurion University of the Negev, Beersheba 8410501, Israel}
%%%%
\author{James Hubble}
\affiliation{Department of Physics, University of Ottawa, 25 Templeton Street, Ottawa, Ontario, K1N 6N5 Canada}
%%%%%
\author{Yingwen Zhang}
\affiliation{National Research Council of Canada, 100 Sussex Drive, Ottawa, ON K1A0R6, Canada}
%%%%%%
\author{Avishy Carmi}
\affiliation{Center for Quantum Information Science and Technology \& Faculty of Engineering Sciences, Ben-Gurion University of the Negev, Beersheba 8410501, Israel}
%%%%%%%
\author{Eliahu Cohen}
\email{eliahu.cohen@biu.ac.il}
\affiliation{Faculty of Engineering and the Institute of Nanotechnology and Advanced Materials, Bar Ilan University, Ramat Gan 5290002, Israel}
%%%%%%%%
\author{Ebrahim Karimi}
\affiliation{Department of Physics, University of Ottawa, 25 Templeton Street, Ottawa, Ontario, K1N 6N5 Canada}
\affiliation{National Research Council of Canada, 100 Sussex Drive, Ottawa, ON K1A0R6, Canada}

\begin{abstract}
Bell inequalities are mathematical constructs that demarcate the boundary between quantum and classical physics. A new class of multiplicative Bell inequalities originating from a volume maximization game (based on products of correlators within bipartite systems) has been recently proposed. For these new Bell parameters, it is relatively easy to find the classical and quantum, i.e. Tsirelson, limits. Here, we experimentally test the Tsirelson bounds of these inequalities using polarisation-entangled photons for different number of measurements ($n$), each party can perform. For $n=2, 3, 4$, we report the experimental violation of local hidden variable theories. In addition, we experimentally compare the results with the parameters obtained from a fully deterministic strategy, and observe the conjectured nature of the ratio. Finally, utilizing the principle of ``relativistic independence'' encapsulating the locality of uncertainty relations, we theoretically derive and experimentally test new richer bounds for both the multiplicative and the additive Bell parameters for $n=2$. Our findings strengthen the correspondence between local and nonlocal correlations, and may pave the way for empirical tests of quantum mechanical bounds with inefficient detection systems.
\end{abstract} %175
\maketitle

\paragraph{Introduction.---} Ever since quantum mechanics was introduced to describe the subatomic world, the foundational aspects, most notably the non-deterministic nature of experimental outcomes, have always been a topic of discussion among physicists and philosophers~\cite{dilip:20}. 
In their seminal work~\cite{EPR}, Einstein, Podolsky and Rosen (EPR) argued for the incompatibility of quantum theory with the idea of local realism. Since then, attempts were made to incorporate extra parameters within the theory, the so called hidden variables,  to ``complete'' the quantum formalism~\cite{hvreview}. 
However, In 1964 John Bell showed that there exist experiments for which any local hidden variable theory must disagree with quantum mechanics about the predicted outcome ~\cite{bell}. This discrepancy is most conveniently illustrated by Bell parameters, i.e., measurable quantities whose values must be bounded to a certain extent in any local hidden variable theory, but can exceed these bounds according to quantum mechanics~\cite{bellnonlocality}. Experiments carried out to test these inequalities~\cite{expttest1,expttest2,expttest3,expttest4} have always vindicated quantum mechanics, thereby showing that local realistic theories do not present an adequate representation of the physical world. Recent new experiments have also significantly progressed towards closing loopholes in a typical Bell experiment, such as freedom-of-choice, fair-sampling, communication (or locality), coincidence and memory loopholes~\cite{freedomchoice,loophole,fair-sampling,Cosmic,Josephson,efficientdetection,BigBell}. 
Several works have attempted to find the extent of Bell parameters involving quantum correlations~\cite{tsirelson,bound1,bound2}. However, finding the classical and quantum (Tsirelson) bounds of these expressions in general remains a challenging task.~\cite{bound1,bound2}. 
Recently, Bell parameters with products of correlators between random variables shared between two parties, namely Alice and Bob, were introduced~\cite{multiplicativebell}. Corresponding to the number of random variables, $n$, measurable by each of the parties, the multiplicative Bell parameter would be proportional to a certain volume in the $n$-dimensional space (this was shown to result from a specific coordination game Alice plays with Bob). For the simplest case where Alice and Bob measure two random variables each, it was also proven that the bound for classical correlations is strictly less than that for quantum correlations. Moreover, these multiplicative Bell inequalities were shown to be more robust to detector inefficiency than the linear ones~\cite{multiplicativebell}. In general, the Tsirelson bound for the multiplicative Bell parameters $\mathcal{B}_n$, corresponding to measurement {\it n}, was shown to be  $\abs{\mathcal{B}_n} \leq n!$. The Tsirelson bounds were derived from the structure of the quantum covariance matrix~\cite{elicovmatrix}, under the assumption of ``relativistic independence''~\cite{relcausality}, mathematically expressing the requirement for locality of uncertainty relations. It was hoped that the nonlinear nature of these inequalities could shed light on the non-polytopic structure of the set of quantum correlations.\newline
Here, we present the results of experiments testing the Tsirelson bounds for these multiplicative inequalities for different $n$ values, and observe their non-intuitive behaviour for large $n$ values. Moreover, we propose and put to test new Tsirelson bounds that are richer than those derived in~\cite{multiplicativebell}.\newline
\begin{figure*}
	\begin{center}
	\includegraphics[width=\linewidth]{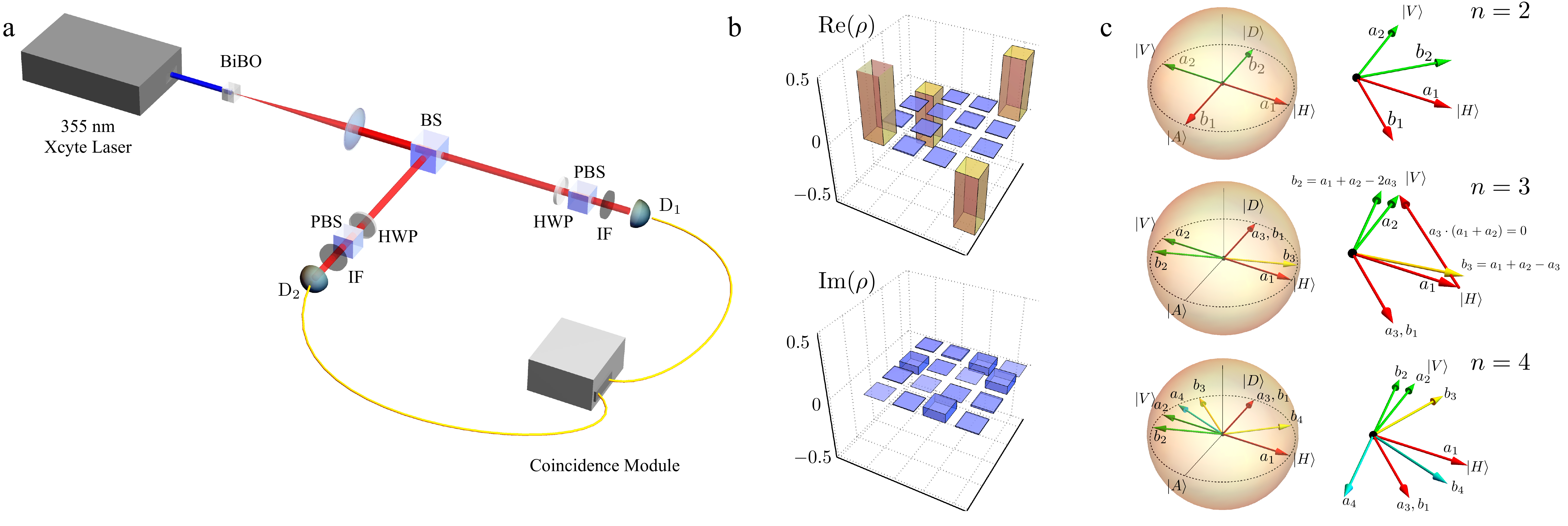}
	\caption[]{Experimental scheme and the chosen strategy to perform measurement of multiplicative Bell parameters. (a) Sketch of the setup used for generating polarisation entangled photons and projecting them onto the states chosen by Alice and Bob's strategies. Entangled photon pairs are generated after pumping paired BiBO (Bismuth Triborate) crystals, and then separated by a 50:50 beamsplitter (BS) into two arms (Alice and Bob). The polarisation measurement stage on each side consists of a half-wave plate (HWP) and polarising beamsplitter (PBS). The photons are filtered by a 710 nm interference bandpass filter (IF), and coupled into single mode fibres, and then detected using single photon avalanche diodes whose signals are sent to a coincidence module from which coincidence events can be observed. (b) Real and imaginary parts of the experimentally reconstructed density matrix of the generated entangled state are shown in $\ket{H}_A\ket{H}_B$, $\ket{H}_A\ket{V}_B$, $\ket{V}_A\ket{H}_B$, and $\ket{V}_A\ket{V}_B$ basis. The generated state fidelity is $\simeq0.977$. (c) The projective measurement strategy, i.e. $\{a_1,a_2, \ldots\}$ and $\{b_1,b_2,\ldots\}$, chosen by Alice and Bob are shown for $n=2$, $n=3$ and $n=4$ on the polarisation Poincar\'e sphere.}
	\label{experimentalsetup}
	\end{center}
\end{figure*} 

\paragraph{Theory.---} Let us consider, a photon pair entangled in the polarisation degree of freedom,
\begin{equation} \label{Bellstate1}
\ket{\psi}= \frac{1}{\sqrt{2}}\left(\ket{H}_{A}\ket{V}_{B}-\ket{V}_{V}\ket{H}_{B}\right),
\end{equation}
where $\ket{H}$ and $\ket{V}$ stand for horizontal and vertical linear polarisation states, and the subscripts $A$ and $B$ represent photon states for Alice and Bob, two spatially separated observers. In the multiplicative Bell scenario, the observers have a different Bell parameter depending upon the number $n$ of different measurements that each can perform. For a general $n$ the Bell parameter is defined as,
\begin{eqnarray}\label{generalbell}
	\mathcal{B}_n &=& \prod_{j=1}^n v_j \cdot c_j\cr
	&=& \left(c_{1n}+\dots+c_{n n}\right) \prod_{j=1}^{n-1} \left(c_{1j}+ \ldots + c_{j j} - j c_{j+1,j} \right),
\end{eqnarray}
where $c_{ij} = c_{a_i b_j} $ is the expectation value of the polarisation measurements along $a_i$ and $b_j$. $c_j$ is the vector comprised of all $ c_{ij} $ for a fixed value of $j$, and $v_j$ is the $j$th column of the matrix:
\begin{equation*}
	V=\begin{bmatrix}
	1 & 1 & \dots & 1 & 1 \\
	-1 & 1 & \dots & 1 & 1 \\
	& -2 & \ddots & \vdots & 1 \\
	&    & \ddots & 1 & \vdots \\
	&    &        & -(n-1) & 1
	\end{bmatrix}.
\end{equation*}
A strategy to select the vectors $a_1 \ldots a_n$, and $b_1 \ldots b_n$, so as to saturate the Tsirelson bounds ($\abs{\mathcal{B}_{n}} \leq n!$) is outlined in~\cite{multiplicativebell}. The multiplicative Bell parameters, $\mathcal{B}_n\triangleq\prod_{j=1}^n v_j \cdot c_j$, can be associated with the additive Bell parameters, $B'_n \triangleq \sum_{j=1}^{n} v_j \cdot c_j$, using the following relation:
\begin{equation}\label{keygg}
	\abs{\mathcal{B}_n} \leq \left(\frac{\mathcal{B}'_n}{n} \right)^n,
\end{equation}
which is a result of the inequality of geometric and arithmetic means, i.e., $\sqrt[n]{\mathcal{B}_n} = \sqrt[n]{ \prod_{j=1}^{n} v_j \cdot c_j } \leq \tfrac{\sum_{j=1}^{n} v_j \cdot c_j}{n}$. The latter expression can be used to find classical upper bounds (not necessarily tight) over $\mathcal{B}_n$, given the classical bounds over $\mathcal{B}'_n$. Using the known fact that additive Bell inequalities are saturated by deterministic strategies~\cite{brunner2014bell}, one may go over all such strategies and find the tight classical bounds for $\mathcal{B}'_n$. The results are as follows:
\begin{eqnarray}
\mathcal{B}'_2 &\leq& 2 \Rightarrow \mathcal{B}_2 \leq 1 ; \cr
\mathcal{B}'_3 &\leq& 5 \Rightarrow \mathcal{B}_3 \leq \left( \frac{5}{3} \right)^3 \approx 4.6 ; \cr
\mathcal{B}'_4 &\leq& 8 \Rightarrow \mathcal{B}_4 \leq 16 . 
\label{classicalbounds}
\end{eqnarray}
$\mathcal{B}'_2$ is the well-known Bell-CHSH parameter~\cite{chsh}. For $n=2$ and $4$, this method yields $\mathcal{B}_2 = 1$ and $\mathcal{B}_4 = 16$, which set the tight classical bounds for these particular values of $n$. This scheme does not yield a useful classical bound for $n>4$.

Since finding the classical limit for the Bell parameter, Eq.~\eqref{generalbell}, is suspected to be difficult in general, an independent and deterministic strategy was proposed for both Alice and Bob ~\cite{multiplicativebell}, and the corresponding limit was calculated. In this strategy, Bob always chooses his random variable to be $+1$, while Alice's choice alternates between $+1$ and $-1$ for all of her variables $A_i$, until $i<i_c$, where $i_c$ is some cutoff number; and for $i>i_c$, she chooses $A_i$ to be $+1$. The correlations take the following values,
\begin{equation}
c_{ij} =
\left\{
	\begin{array}{ll}
		(-1)^i  &  i \leq i_c \\
		1 &  i > i_c
	\end{array}
\right. .
\end{equation}
The value for the cutoff $i_c$ is taken such that it maximizes the value for the Bell parameter. Analytically, the value $\mathcal{B}_n$ obtains for this fully deterministic strategy, $\text{FD}_n$, can be explicitly calculated, and its value is,
\begin{equation}
    \text{FD}_n= 2^{i_c}\,\left[\left(\frac{i_c}{2}\right)!\right]^2\,\left(n-i_c\right)\, i_{c}^{(n-i_c-1)}.
    \label{fdn}
\end{equation}
The maximal values obtained from the fully deterministic strategy for $n=2$ and $4$, coincide with the classical bounds obtained in Eq.~\eqref{classicalbounds}.
The ratio of fully deterministic strategy and Tsirelson bound, i.e. $\text{FD}_n/n!$, approaches $\sqrt{{\pi}/{2e}}$ as $n \to \infty$.

\begin{figure}
	\includegraphics[width=\linewidth]{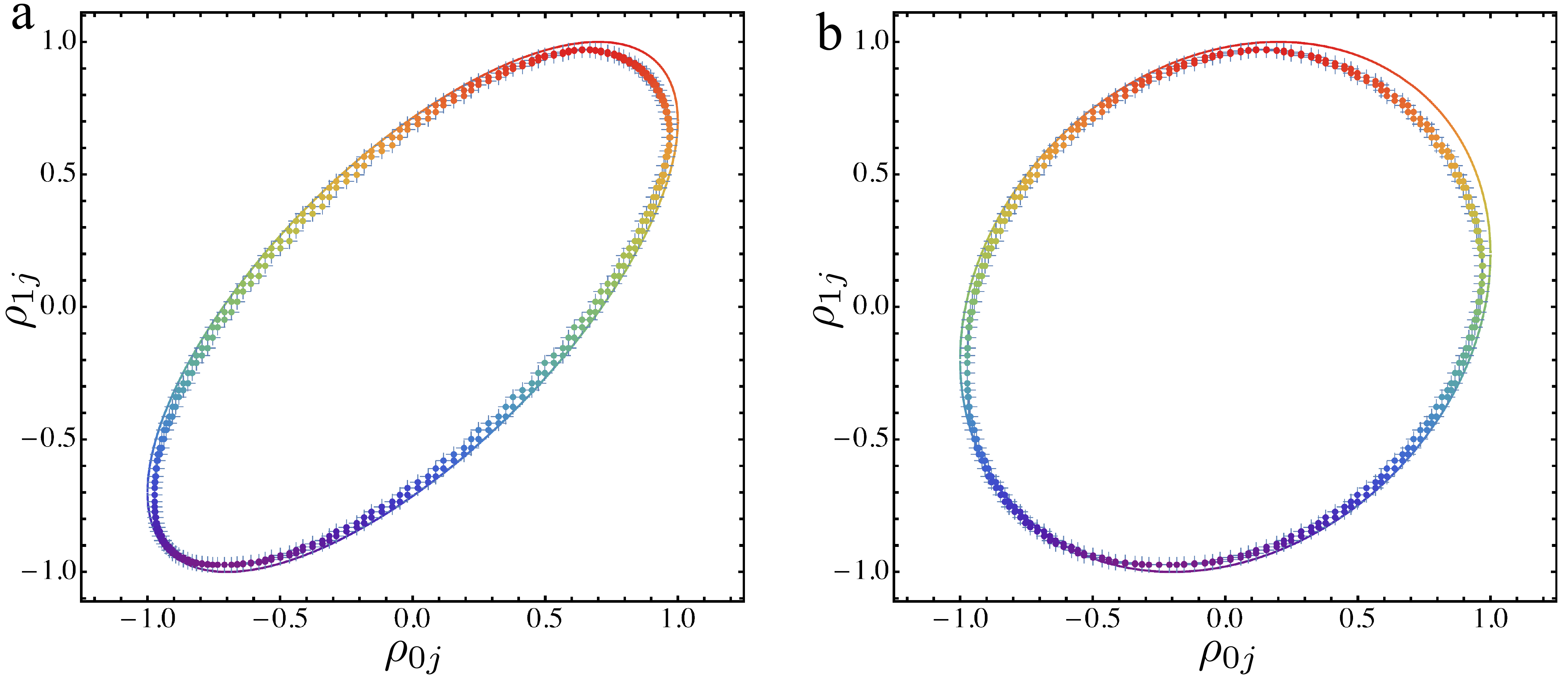}
	\caption[]{Distribution of the correlation vectors for (a) $\eta_A = 0.7$ and (b) $\eta_A=0.2$. The solid line indicates the region within which the correlation vectors should fall, and the points are experimentally measured vectors.}
	\label{ellipse}
\end{figure}

\paragraph{Experiment.---} Paired 0.5-mm-thick type-I bismuth triborate crystals (BiBO), one rotated by 90-degrees with respect to the other, are pumped by a quasi-continuous wave 100 mW, 355 nm beam to generate photon pairs (signal and idler) via spontaneous parametric downconversion (SPDC) at a degenerate wavelength of 710~nm~\cite{Kwiat} -- see Figure~\ref{experimentalsetup}(a). A half-wave plate, placed before the paired crystals, is used to tailor the photon pair state by controlling the pump beam polarisation state. Due to indistinguishably, the generated photon pairs are entangled in the polarisation degree of freedom. The polarisation of the pump beam and the orientation of the crystal was tuned so as to obtain the state, $\ket{\psi}_\text{SPDC}=\tfrac{1}{\sqrt{2}}\big(\ket{H}_A\ket{H}_B- \ket{V}_A\ket{V}_B\big)$. A half-wave plate was placed in one of the arms, say Bob, so that the horizontal and vertical polarisation states are exchanged and state in Eq.~\eqref{Bellstate1} was obtained. The 355 nm pump beam is afterwards filtered out with a long pass filter. The photon pairs are separated by a 50:50 beamsplitter (BS), and the photons travel along two different arms. To select two diametrically opposite regions of the SPDC cone, we place a pair of irises on each arm such that they center on the required region. In each arm projective measurement of the polarisation state is performed by a combination of a quarter-wave plate, a half-wave plate, and a polarizing beam splitter (PBS). Bandpass filters of $(710 \pm 5)$ nm are placed before 20$\times$ objectives so that only degenerate photon pairs are coupled into single mode fibers of core diameter $5~\mu$m. Then, the photons are detected via a pair of single photon avalanche diode (SPAD) detectors (Excelitas SPCM-AQRH-14-FC), and are finally  counted via a time-correlated single photon counting system. To achieve a maximally entangled state of Eq.~\eqref{Bellstate1}, the angle of the waveplate before the crystal and the orientation of the crystal was appropriately adjusted. The coincidence rates depending upon the difference in the two half-wave plate orientations, ranged from $5220~s^{-1}$ to $34~s^{-1}$. The visibility in $H/V$ basis and $\pm 45^\circ$ linear polarisation states were $(98.2 \pm 0.5)\%$, and $(97.3 \pm 0.5)\%$, respectively. Real and imaginary parts of the generated photons' density matrix are shown in Figure~\ref{experimentalsetup}(b). The fidelity of the generated state from the reconstructed density matrix was measured to be $\simeq0.977$, confirming the high quality of the entangled source. Figure ~\ref{experimentalsetup}(c) illustrates the set of vectors used in projecting Alice and Bob's quantum states, i.e. $\{a_1,a_2,a_3, \ldots\}$ and $\{b_1,b_2,b_3, \ldots\}$, on the polarisation Poincar\'e sphere. 
\begin{table}[ht!]
	\caption[]{\textbf{Classical ($\mathcal{B}_n^\text{Classical}$), Fully deterministic $\text{FD}_n$, and Tsirelson ($\mathcal{B}_n^\text{Quantum}$) bounds for the multiplicative Bell parameters, and the experimentally ($\mathcal{B}_n^\text{Experiment}$) measured values.}}
	\begin{tabular}{p{1.0 cm} p{1.5 cm} p{1.5 cm}  p{1.8 cm}  p{1.8 cm} }
	\hline \hline
		n & $\mathcal{B}_n^\text{Classical}$ &
		$\text{FD}_n$ & $\mathcal{B}_n^\text{Experiment}$ &	$\mathcal{B}_n^\text{Quantum}\,(n!)$ \\ \hline
		$2$ & $1$ & $1$ & $1.88 \pm 0.05 $ & $2$  \\ 
		$3$ & $4-4.6$  & 4 & $5.85 \pm 0.31$  & $6$   \\ 
		$4$ & $16$ & 16 & $23.3 \pm 1.4 $ & $24$    \\ 
		$5$ & N.A. & 64 & $115 \pm 9$ & 120 \\ 
		$6$ & N.A. & 512 & $687 \pm 59$ & 720 \\ 
		$7$ & N.A. & 3072 & $4655 \pm 374$ & 5040 \\ [1 ex]
		\hline
	\end{tabular}
	\label{table1}
\end{table}
The number of photon counts $N$ detected in coincidence along $a$ and $b$ by Alice and Bob's detectors was recorded, from which one can obtain the expectation value of the measurement via $c_{a\,b} ={\left(N_{++}-N_{+-}-N_{-+}+N_{--}\right)}/{\left(N_{++}+N_{+-}+N_{-+}+N_{--}\right)}$, such that $N_{+-}$, for instance, refers to the joint measurement where Alice and Bob set their apparatus to measure the state along the positive and negative direction of $a$ and $b$, respectively.  Using this approach, as a verification, the CHSH parameter for our single photon source is measured and found to be $ 2.748 \pm 0.026 \leq 2\sqrt{2}$, which lies beyond the classical limit of $2$.
The Bell parameter for $n=2$ is $\mathcal{B}_2 = \abs{(c_{12}+c_{22})(c_{12}-c_{21})}$. The classical and quantum limits of $\mathcal{B}_2$ are, respectively, $\mathcal{B}_2^\text{Classical}\leq1$ and $ \mathcal{B}_2^\text{Quantum}\leq 2! $. The value for $\mathcal{B}_2$ as calculated from our experiment was $\mathcal{B}_2 = 1.88\pm 0.05 $, which is beyond the classical limit $1$. For $n=3$ the Bell parameter is given by, $\mathcal{B}_3 = \abs{(c_{13}+c_{23}+c_{33})(c_{11}-c_{21})(c_{12}+c_{22}-2c_{32})}$. The classical limit for $\mathcal{B}_3$ lies somewhere between $4$ and $4.6$. Experimentally the observed value for $\mathcal{B}_3$ is $5.85 \pm 0.31$. Experimentally measured values for the Bell parameters up to $n=7$ with the Tsirelson bounds and the classical limits (where applicable) are shown in Table \ref{table1}.\newline
\begin{figure}
	\includegraphics[width=\linewidth]{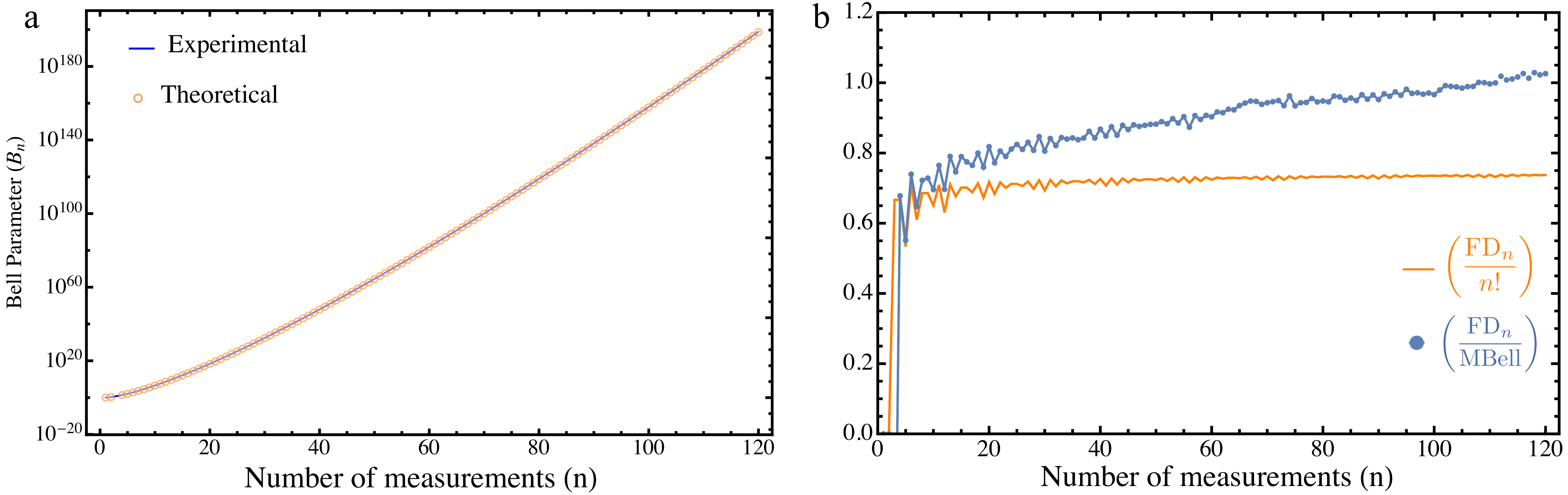}
	\caption[]{Experimentally calculated multiplicative Bell parameters (MBell). (a) Logarithmic plot of the experimentally calculated Bell parameters (blue), and theoretical quantum limit of the Bell parameter, i.e. n!, (red). (b) Ratio of the parameters generated from the fully deterministic strategy $\text{FD}_n$ taken with the Tsirelson bound, $n!$ (red), and with the experimentally observed Bell parameter MBell (blue).}
	\label{logplotofbell}
\end{figure} 
The experimentally measured Bell parameters for higher values of $n$ are plotted in Figure \ref{logplotofbell}(a). The primary contribution to the uncertainties in the Bell parameter is associated with the rotation of the half-wave plates in the detection state. Since the coincidence counts were taken by rotating motor controlled waveplates at intervals of $1^\circ$, the corresponding maximum and the average uncertainties in the expectation values are $0.0698$ and $0.044$, respectively. The uncertainty associated with photon counting  from Poissonian statistics contributes to maximum uncertainty of $0.003$. Thus the uncertainties, are dominated by the rotation of the waveplates, which is extremely small, and not visible in Figure \ref{logplotofbell}(a). The obtained results are close to the corresponding Tsirelson bounds. In order to quantify the results, we plot the ratio between the maximal quantum bound and the observed parameters, see Figure \ref{logplotofbell}(b). Most of the obtained results are within eighty percent of the maximal value.
\smallskip
As a comparison with the fully deterministic strategy~\eqref{fdn}, the theoretical ($\text{FD}_n/n!$) and the obtained experimental ($\text{FD}_n/\text{MBell}$) ratios are shown in Figure~\ref{logplotofbell}(b) for values of $n$ up to $255$. As the experimentally observed Bell parameters are less than the actual ones the experimentally calculated values for the ratio are slightly larger than the theoretical ones.
\begin{figure}
	\includegraphics[width=\linewidth]{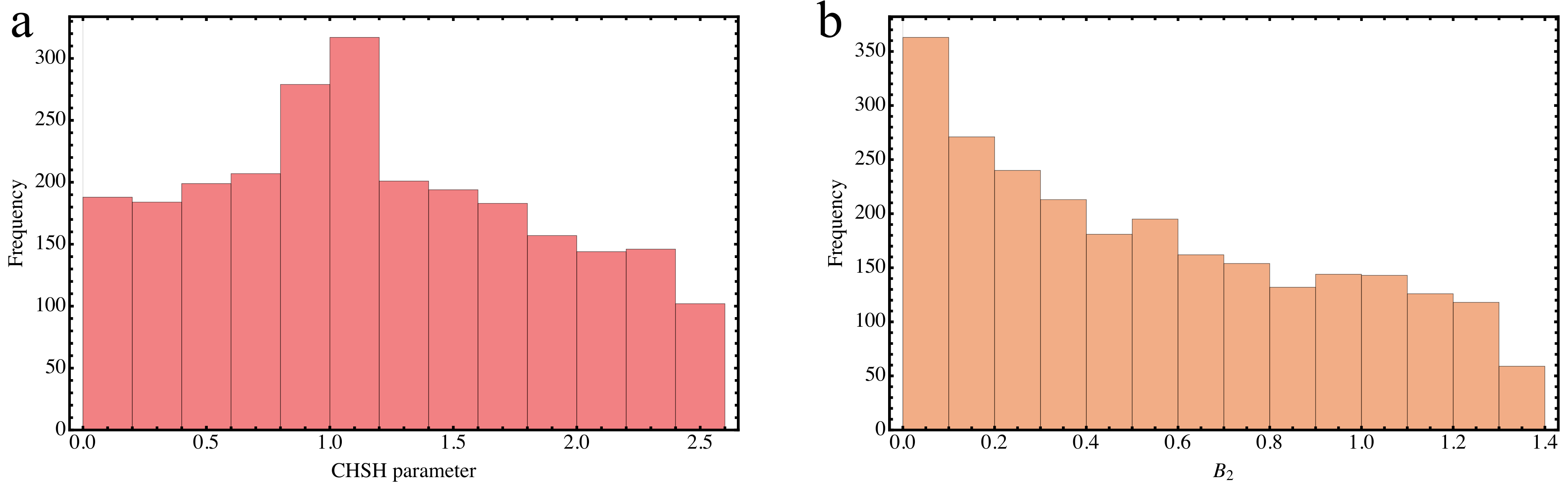}
	\caption[]{Distribution of both (a) CHSH parameters (upper limit = 2.618) as well as (b) Bell parameters $\mathcal{B}_2$ (upper limit =1.428), for $\eta_A = 0.7$.}
	\label{upperbounds}
\end{figure}
\paragraph{Locality of uncertainty and richer quantum bounds.---} Recently it was proposed~\cite{relcausality} that the locality of uncertainty, i.e. the requirement that local uncertainties are independent of the measurement choices of any other parties, can give rise to both known and hitherto unnoticed bounds on nonlocal correlations in any statistically meaningful theory. For two parties sharing a Bell state, richer bounds for the CHSH parameter and the multiplicative Bell parameter can be obtained in terms of the local correlations. Let us define the local correlation, say on Alice's side, 
\begin{equation}
    \eta_A = \frac{\braket{A_iA_j}-\braket{A_i}\braket{A_j}}{\Delta_{A_i} \Delta_{A_j}},
\end{equation}
where $A_i$, and $A_j$ are two local observables for Alice, and $\Delta^2_{A_{k}}= \braket{A_{k}^2} - \braket{A_{k}}^2$. In the Supplementary Material we show that these local correlations on Alice's side give rise to new bounds for the standard CHSH and the multiplicative Bell parameters, which respectively are $\text{CHSH} \leq \sqrt{2}\left(\sqrt{1+\eta_A}+\sqrt{1-\eta_A}\right)$ and $\mathcal{B}_{2} \leq 1 + \sqrt{1-\eta_A^2}$. For the special case of maximally entangled states, as the one considered here, the multiplicative Bell parameter is more tightly bounded as follows, $\mathcal{B}_{2} \leq 2 \sqrt{1-\eta_A^2}$ (for derivation  please see the Supplemental Material). For both bounds, the local correlation given by $\eta_A$ determines the extent of the nonlocal correlations on the right hand side, i.e. upper bound. These inequalities are derived based on Alice's local correlations, but can be similarly derived for Bob's, and hence an even tighter bound is given by their minimum. To experimentally infer these richer bounds on correlations, we selected two vectors $a_1$, and $a_2$ on Alice's side that correspond to a particular value of $\eta_A$. Then we randomly selected two vectors on Bob's side, and calculated the CHSH and Bell parameters. In Figures~\ref{upperbounds}(a) and (b), we show the results for both parameters when $\eta_A =0.7 $. The observed values all fall within the bound as predicted by $\text{CHSH} \leq \sqrt{2}\left(\sqrt{1+\eta_A}+\sqrt{1-\eta_A}\right)$ and $\mathcal{B}_{2} \leq 1 + \sqrt{1-\eta_A^2}$. The local correlations, for example, $\eta_A$, also put restrictions on the two particle correlations, which can be defined as,
\begin{equation}
    \rho_{i j}=\frac{\braket{A_i B_j}-\braket{A_i}\braket{B_j}}{\Delta_{A_i}\Delta_{B_j}},
\end{equation}.
Geometrically, for a particular $\eta_A$ the correlation vectors $\left(\rho_{0j},\rho_{1j}\right)$, $j= 0, 1$, lie on the ellipse whose major and minor axes are related to $\eta_A$ as,
\begin{equation}
    e_{j}=\sqrt{\left(\frac{1 \pm (-1)^j \abs{\eta_A}}{\sqrt{2}}\right)}
    \begin{bmatrix}
    1 \\
    (-1)^j
    \end{bmatrix}.
\end{equation}
Similar relation holds for the vectors $\left(\rho_{i0},\rho_{i1}\right)$, $i= 0, 1$, defined by the local correlation $\eta_B$, on Bob's side. Figure ~\ref{ellipse} shows the ellipses corresponding to particular values of $\eta_A = 0.7$, and $\eta_A = 0.2$, along with the experimentally measured correlation vectors.

\paragraph{Conclusion.---}
We derived new classical and quantum bounds for the multiplicative Bell inequalities and experimentally tested them. Our theoretical results mainly stem from the principle of relativistic independence \cite{relcausality} and hence emphasize the significance of correlations.The experimental results show that the multiplicative Bell parameters go beyond their classical limits, thus again falsifying local realism. Additionally, they approach the Tsirelson bound, the upper limit derived from the quantum covariance matrix, and the ``relativistic independence'' assumption. We were also able to experimentally observe the tighter theoretical bounds on the CHSH and the Bell parameter, $\mathcal{B}_2$, lending support to the claim that quantum correlations arise from the locality of uncertainty relations \cite{relcausality}. An additional practical merit of multiplicative Bell inequalities, is that they elevate the detector efficiency requirement~\cite{multiplicativebell} for the test of hidden variable theories. It is challenging for Bell experiments to reach high detection efficiencies, but in principle, multiplicative Bell inequalities, like those we examined here, provide a simple circumvention of the detector efficiency loophole prevalent in Bell experiments.
\newline
The authors would like to thank Alessio D'Erico for fruitful discussions. This work was supported by Canada Research Chairs (CRC), Canada Foundation for Innovation (CFI), and Canada First Excellence Research Fund (CFREF). E.C. acknowledges support from the Israel Innovation Authority under project 70002 and from the Quantum Science and Technology Program of the Israeli Council of Higher Education.

\clearpage
\renewcommand{\figurename}{\textbf{Supplementary Figure}}
\setcounter{figure}{0} \renewcommand{\thefigure}{\textbf{{\arabic{figure}}}}
\setcounter{table}{0} \renewcommand{\thetable}{S\arabic{table}}
\setcounter{section}{0} \renewcommand{\thesection}{S\arabic{section}}
\setcounter{equation}{0} \renewcommand{\theequation}{S\arabic{equation}}
\onecolumngrid

\section*{Supplementary Material for Experimental tests of Multiplicative Bell Inequalities}\label{SM}

Here we derive the richer CHSH and Multiplicative Bell inequalities for $n=2$.
Let us define the Pearson correlations in a standard Bell-\text{CHSH} experiment as,

\begin{equation}
\rho_{ij} \ass \frac{\langle A_i B_j \rangle - \langle A_i \rangle \langle B_j \rangle} {\cv{A_i} \cv{B_j}},
\end{equation}
where the variances are $\cv{A_i}^2 = \langle A_i^2 \rangle - \langle A_i
\rangle^2$ and $\cv{B_j}^2 = \langle B_j^2 \rangle - \langle B_j
\rangle^2$. In any theory satisfying the generalized uncertainty relation presented in
\cite{relcausality}, the principle of relativistic causality implies
\begin{subequations}
\label{eq:r1}
\begin{equation}
\begin{bmatrix}
1 & \eta_A \\
\eta_A^{*} & 1
\end{bmatrix} \succeq
\begin{bmatrix}
\rho_{0j} \\
\rho_{1j}
\end{bmatrix}
\begin{bmatrix}
\rho_{0j} &
\rho_{1j}
\end{bmatrix}
\end{equation}
\begin{equation}
\begin{bmatrix}
1 & \eta_B \\
\eta_B^{*} & 1
\end{bmatrix} \succeq
\begin{bmatrix}
\rho_{i0} \\
\rho_{i1}
\end{bmatrix}
\begin{bmatrix}
\rho_{i0} &
\rho_{i1}
\end{bmatrix}
\end{equation}
\end{subequations}
for $i,j \in \{0,1\}$, where $\eta_A$ and $\eta_B$ are two complex numbers satisfying $\abs{\eta_A} \leq 1$, $\abs{\eta_B} \leq
1$, and $\eta^{*}$ denotes the complex conjugate of
$\eta$. Quantum mechanics satisfies both relativistic causality and the generalized uncertainty
relations~\cite{relcausality}, for
\begin{equation}
\label{eq:g1}
\eta_A \ass \frac{\langle A_0 A_1 \rangle - \langle A_0 \rangle \langle
  A_1 \rangle}{\cv{A_0} \cv{A_1}}, \; \; \;
\eta_B \ass \frac{\langle B_0 B_1 \rangle - \langle B_0 \rangle \langle
  B_1 \rangle}{\cv{B_0} \cv{B_1}}.
\end{equation}

It was also shown in~\cite{relcausality} that following Tsirelson-like bounds stem from  \eqref{eq:r1} and \eqref{eq:g1},:
\begin{subequations}
\label{eq:ts1}
\begin{equation}
\abs{\text{CHSH}} \leq \min \left \{ \sqrt{2} \left( \sqrt{1 + \mathrm{Re}(\eta_A)} +
\sqrt{1 - \mathrm{Re}(\eta_A)} \right), \; \; 2 \sqrt{2} \sqrt{1
- \mathrm{Im}(\eta_A)^2} \right \} \leq 2 \sqrt{2}
\end{equation}
\begin{equation}
\abs{\text{CHSH}} \leq \min \left \{ \sqrt{2} \left( \sqrt{1 + \mathrm{Re}(\eta_B)} +
\sqrt{1 - \mathrm{Re}(\eta_B)} \right), \; \; 2 \sqrt{2} \sqrt{1
- \mathrm{Im}(\eta_B)^2} \right \} \leq 2 \sqrt{2}
\end{equation}
\end{subequations}
where $\text{CHSH} \ass \rho_{00} + \rho_{10} + \rho_{01} -
\rho_{11}$ is the Bell-\text{CHSH} parameter.

For real $\eta$, Eq. \eqref{eq:ts1} transforms to,
\begin{subequations}
\begin{equation}
\label{eq:ts}
\abs{\text{CHSH}} \leq  \sqrt{2} \left( \sqrt{1 + \eta_A} +
\sqrt{1 - \eta_A} \right) \leq 2 \sqrt{2}
\end{equation}
\begin{equation}
\abs{\text{CHSH}} \leq  \sqrt{2} \left( \sqrt{1 + \eta_B} +
\sqrt{1 - \eta_B} \right) \leq 2 \sqrt{2}
\end{equation}
\end{subequations}
For the multiplicative Bell parameter, $\mathcal{B}_2$, using the inequality of geometric and arithmetic means and using Eq. \ref{eq:ts},
\begin{eqnarray}
    \abs{\mathcal{B}_2} &=& \abs{( \rho_{00}+\rho_{10})(\rho_{01}-\rho_{11})} \leq \left( \frac{\rho_{00}+\rho_{10}+\rho_{01}-\rho_{11}}{2} \right)^2 \\
    \abs{\mathcal{B}_2} &\leq& \frac{1}{2} \left(\sqrt{1-\eta_A}+\sqrt{1+\eta_A}\right)^2 \nonumber\\
    &=& 1 + \sqrt{1-\eta_A^2}. \label{mulbound}
\end{eqnarray}

In~\cite{elicovmatrix}, it was also proven that,
%
%\begin{eqnarray}
%    \Abs{\braket{X_i}\pm\braket{X_j}} \leq \sqrt{2(1 \pm M_{ij})}
%\end{eqnarray},
%

\begin{eqnarray}
   &\abs{\rho_{00}+ \rho_{01}} \leq \sqrt{2(1 + d)} \label{m1} \\
   &\abs{ \rho_{10}- \rho_{11}} \leq \sqrt{2(1 - d)} \label{m2},
\end{eqnarray}
where,
$d=\braket{\{A_0,A_1\}}/2$, and $\{A_0,A_1\}$ is the anti commutator of the  local operators $A_0$, and $A_1$.
For a maximally entangled state like the singlet state,
the expectation values of the local operators are zero, and 
$d=\eta_A$.
From Eq. \ref{m1}, and Eq. \ref{m2},
\begin{equation}
    \abs{\mathcal{B}_2}=\abs{(\rho_{00}+ \rho_{01})(\rho_{10}- \rho_{11})}\leq 2\sqrt{1-\eta_A^2}.
\end{equation}

The role of local correlations in determining the non local,
Alice-Bob correlations is evident in all these
characterizations. Using the above definitions, one may also recognize
\begin{equation}
\label{eq:unc}
\abs{\eta_A}^2 = \mathrm{Re}(\eta_A)^2 + \mathrm{Im}(\eta_A)^2 \leq 1,
\; \; \; \abs{\eta_B}^2 = \mathrm{Re}(\eta_B)^2 +
\mathrm{Im}(\eta_B)^2 \leq 1
\end{equation}
as the Schr\"{o}dinger uncertainty relations of Alice's $A_0$ and
$A_1$, and of Bob's $B_0$ and $B_1$.

\end{document}